\documentclass[a4paper]{article}

\usepackage{INTERSPEECH2021}
\usepackage{amsfonts}
\usepackage{amsbsy}
\usepackage{amsmath}
\usepackage{microtype}
\usepackage[ruled,vlined]{algorithm2e}
\usepackage{xcolor}
\usepackage{makecell}
\usepackage{multirow}
\usepackage{adjustbox}
\usepackage{rotating}
\usepackage{multirow}
\usepackage{hhline}
\usepackage{graphicx}
\usepackage{hyperref}

\title{Flavored Tacotron: Conditional Learning for Prosodic-linguistic Features}

\name{Mahsa Elyasi, Gaurav Bharaj}
\address{
  AI Foundation, USA
}
\email{{\{mahsa, gaurav}\}@aifoundation.com}

\begin{document}
\maketitle
\begin{abstract}
Neural sequence-to-sequence text-to-speech synthesis~(TTS), such as Tacotron-2, transforms text into high-quality speech. However, generating speech with natural prosody still remains a challenge. Yasuda et. al.~\cite{yasuda2020investigation} show that unlike natural speech, Tacotron-2’s encoder doesn't fully represent prosodic features (e.g. syllable stress in English) from characters, and result in flat fundamental frequency variations. 

In this work, we propose a novel carefully designed strategy for conditioning Tacotron-2 on two fundamental prosodic features in English -- stress syllable and pitch accent, that help achieve more natural prosody. To this end, we use of a classifier to learn these features in an end-to-end fashion, and apply feature conditioning at three parts of Tacotron-2’s Text-To-Mel Spectrogram: pre-encoder, post-encoder, and intra-decoder. Further, we show that jointly conditioned features at pre-encoder and intra-decoder stages result in prosodically natural synthesized speech (vs. Tacotron-2), and allows the model to produce speech with more accurate pitch accent and stress patterns.

Quantitative evaluations show that our formulation achieves higher fundamental frequency contour correlation, and lower Mel Cepstral Distortion measure between synthesized and natural speech. And subjective evaluation shows that the proposed method's Mean Opinion Score of 4.14 fairs higher than baseline Tacotron-2, 3.91, when compared against natural speech (LJSpeech corpus), 4.28.


\end{abstract}

\noindent\textbf{Index Terms}: Text-to-speech synthesis, English prosody, Prosodic feature, Pitch accent, End-to-end learning, Tacotron 2

\section{Introduction}
\label{sec:intro}

Text-To-Speech~(TTS), a sequence-to-sequence problem, aims to synthesize intelligible and natural sounding speech from input text. Generally, TTS approaches can be categorized as: statistical parametric speech synthesis~(SPSS) and Neural sequence-to-sequence TTS. SPSS typically consists of several domain-specific modules that require feature engineering: a text analyzer to convert input text into linguistic and prosodic features (front-end), a duration model to predict phoneme duration, an acoustic model (back-end) to generate fundamental frequency contour~($F_{0}$) and spectrum, and a vocoder to synthesize speech from these acoustic features. Recent methods use neural sequence-to-sequence techniques that represent these internal modules as a single neural module. Here, output speech is directly inferred from input text; this technique is ideally called end-to-end TTS~\footnote{We refer to them as neural sequence-to-sequence methods, since in-practice, they are not fully end-to-end, where separate vocoder and grapheme-to-phoneme methods are used for quality improvement.}.

The main advantage of a neural sequence-to-sequence TTS is that it does not require explicit feature engineering or feature extraction. An encoder-decoder architecture transforms input text to an intermediate representation (Mel-Spectrogram) followed by a vocoder that infers speech waveform directly from Mel-Spectrogram. The goal of encoder is to extract robust linguistic and prosodic features from input text. For example, encoder architecture in neural TTS method, Tacotron~2~\cite{shen2018natural}, is inspired from work done in the field of machine translation, that contracts a space to transform input text into a cross-language linguistic representation. Such an encoder architecture does not fully represent prosodic-linguistic features, since it is not designed to explicitly assimilate prosodic information.

This inability to encode text-based prosodic-linguistic features is a drawback of most neural sequence-to-sequence TTS methods, when compared with SPSS (e.g. Merlin~\cite{wu2016merlin}) that uses a front-end text processors (e.g. Festival~\cite{black1998festival}) to extract linguistic and prosodic-linguistic features.  One way to bridge the gap between SSPS and neural sequence-to-sequence TTS is to enrich the input sequence with explicit linguistic features. While, adding a complete set of engineered features is counterintuitive to the neural sequence-to-sequence TTS \cite{watts2019improvements}, we note that adding some structure can help guide neural TTS towards better quality. Other works~\cite{liu2019cross, liu2020multi, fujimoto2019impacts, yasuda2019investigation, luong2018investigating} show that use of pitch accent (Japanese and English), tone (Chinese) and stress syllables (Japanese, Chinese and English) as additional inputs can help improve subjective and quantitative evaluations. To the best of your knowledge, conditioning pitch accent and stress syllable has not been shown for English Tacotron~2, neither is applying it in a supervised fashion or at other parts of Tacotron~2 rather than input. In this work, we present a method to use a minimal set of features that help improve overall quality of synthesized speech. To summarize, our contributions include: 
\begin{itemize}
\item A carefully designed strategy that conditions learnable prosodic features in the Tacotron~2’s Text-To-Mel Spectrogram~(TTM).
\item Use of two fundamental prosodically related lexical features in English language -- stress syllable and pitch accent.
\item Improved overall prosodic quality of generated speech, quantitative and subjective evaluations.
\end{itemize}

In Section~\ref{sec:rw}, we summarize the details of neural sequence-to-sequence TTS and the usage of prosodic-linguistic features. In Section~\ref{sec:method}, we propose a new formulation for conditioning Tacotron~2 on these features – stress syllable and pitch accent. In Section~\ref{sec:results}, we present the our quantitative and subjective evaluations. Finally, in Section \ref{sec:conclusion}, we discuss our findings and the future works.

\section{Related Works}
\label{sec:rw}
\begin{figure*}[hbt!]
  \includegraphics[width=\linewidth]{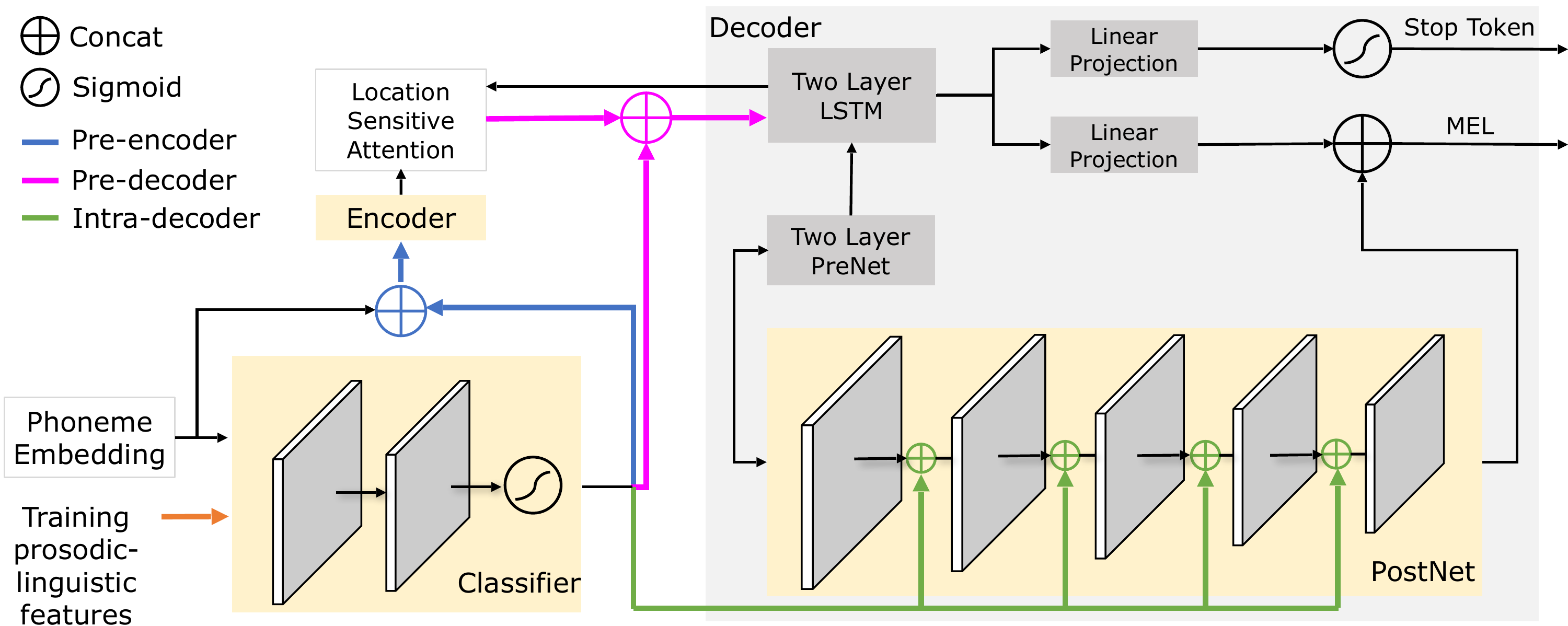}
   \caption{Proposed  structural  changes applied to Tacotron 2’s TTM for conditioning on prosodic-linguistic features. The pre-encoder, pre-decoder, and intra-decoder methods are differentiated through blue, magenta, and green arrows. The red arrows only applied during the training}
   \label{fig:networkstructure}
\end{figure*}
Comprehensive linguistic representation is feasible when TTS's encoder and the training data are sufficiently rich in complexity and large, respectively~\cite{yasuda2020investigation}. Increasing the complexity of the encoder requires an increase in model size, and resultant training iterations. Similarly, creating a dataset that matches the word-coverage of a lexicon dictionary is challenging. For example, even large-scale data such as LibriTTS~\cite{zen2019libritts} cannot offer such a comprehensive word-coverage~\cite{taylor2019analysis}. Thus, several studies attempt to use linguistically richer inputs. Regardless of language, usage of phoneme input helps generate better results than character input~\cite{yasuda2020investigation}, in neural sequence-to-sequence TTS~(e.g., Tacotron~2).    

Prosodic-linguist features, in addition to the phoneme inputs, is essential in tonal~(such as Chinese) and pitch accent languages~(such as Japanese) due to failure of neural sequence-to-sequence TTS systems in generating intelligible speech from character only inputs. Thus, a combination of tonal types and stress for Chinese, and pitch accent types and stress for Japanese are often used as an additional prosodic-linguist input features.

Two ways to add these features to the phoneme input include: augmenting~\cite{liu2020tone, suni2020prosodic} and conditioning~\cite{liu2020multi, luong2018investigating}. When augmenting, the length of phoneme set is increased with respect to the number of added features, while dimension of the phoneme embedding remains unchanged. When conditioning, the length of phoneme set remains unchanged, and dimension of the phoneme embedding is increased with respect to the number of added features~\footnote{For example, consider a language with $n$ voiced and $m$ unvoiced phonemes and a phoneme embedding mechanism with $k$ dimension. For augmenting the syllable stress feature, the length of the phoneme set will increase to $(2)n + m$, while for conditioning the same features, size of the phoneme embedding will increase to $k + 1$.}. 

Even though high quality synthesized speech using Tacotron~2, from character input has been reported for English language~\cite{shen2018natural}, many studies prefer to use phoneme input rather than the character input due to presence of mispronunciation and inaccurate stress levels when the character input is used. One reason for this preference is that the CNN-based encoder from Tacotron~2 is more lightweight when compared to CBHG encoder, Tacotron~\cite{wang2017tacotron} which makes it challenging to learn the the disambiguation between underlying character pronunciation and stress syllable patterns. In a comprehensive study, Yasuda et. al.~\cite{yasuda2020investigation, yasuda2019investigation} have investigated the effects of linguistic features in Tacotron based synthesis in comparison with several SPSS systems for two languages English and Japanese. They show that using phoneme input (augmented with stress syllable) significantly improves the naturalness of synthesized speech. In analysis of the synthesized speech with MOS lower than 2.5, unnatural prosody was established as the main cause. They also report that English Tacotron based systems generate flatter $F_{0}$ contour than SSPS systems and natural speech, and result in unnatural prosody and lower MOS measure. Similarly, Shen et. al.~\cite{shen2018natural} note that unnatural prosody (specifically unnatural pitch accent) as the main artifact in an analysis of English sentences.

Augmenting the phoneme inputs with stress syllables is a common practice in training English Tacotron~2. Conditioning the stress syllable features into the phoneme embedding vector is used to construct a multi-lingual model~\cite{liu2020multi, liu2019cross}. Suni et. al.~\cite{suni2020prosodic} uses the augmentation method to add three prosodic-linguistic features: stress syllable~(unstressed and stressed), pitch accent~(unaccented, accented, and emphasized), and phrase boundary~(no phrase, minor phrase and major phrase). The stress syllable are added into the voiced phonemes (resulting two symbols per each), while pitch accent and phrase boundary are encoded into nine symbols that are each added before prominent word or a word followed by a phrase boundary. Our method differs from \cite{suni2020prosodic} in the following ways: 1) they use raw features (extracted from speech) during training and test, while we learn prediction of these features (extracted from text) during the training. 2) they augment the features in to the input of the Tacotron~2, while we condition Tacotron-2 on these features at three different modules. 
\section{Method}
\label{sec:method}

We first introduce our baseline, then propose a carefully designed strategy for conditional learning of prosodic features -- stress syllable and pitch accent, on the baseline. And in Section~\ref{subsec:objective}, we justify our choice for the proposed conditioning strategy.

\subsection{Baseline}
We use Tacotron~2's TTM to synthesize a Mel Spectrogram from input text, and we use Parallel-Wavegan's  Vocoder~\cite{yamamoto2020parallel} to synthesize waveforms from synthesized Mel Spectrogram features. We refer to the combination of Tacotron~2's TTM and Parallel-Wavgan's Vocoder as our baseline (see Table~\ref{tbl:networkstructure} for baseline's details).

\subsection{Proposed method}
We note that in computer vision, it has been shown that adding structured noise before every CNN modules of the network results in more accurate image generation~\cite{karras2019style}. Further, coordinated CNN~\cite{liu2018intriguing} shows that by conditioning a CNN module with the input coordinates leads to more accurate prediction of object coordination. In speech synthesis, it has been shown that conditioning/augmenting linguistic features into the input improves the speech naturalness. Also, conditioning speaker IDs before Tacotron's decoder is commonly used for training a multi-speaker TTS system. We take inspiration from such approaches, and propose a novel structured conditioning strategy that results in richer local variation in $F_{0}$ contour.

Figure~\ref{fig:networkstructure}, illustrates the proposed structural changes to Tacotron~2's TTM for conditioning on prosodic-linguistic features. We use a classifier that take the output of the phoneme embedding as input and predict two dimensional binary vector as an output. This classifier consists of two layers bidirectional LSTM followed by a fully connected network and Sigmoid activation. Bi-LSTM followed by Conditional Random Field layer is commonly used architecture for sequential labeling tasks. Since, only two features need be predicted, we do not need a heavier structure after the Bi-LSTM layers. Therefore, we use a fully connected network with a Sigmoid activation.

We then use the learnt binary vector to condition at three stages of Tacotron~2's TTM : 
\begin{itemize}
    
\item Pre-encoder: where output of phoneme embedding and classifier are concatenated and encoded by the encoder. \item Pre-decoder: where the output of attention and classifier are concatenated and passed to the decoder. \item Intra-decoder: where the output of each CNN model in post-net and classifier are concatenated and passed to the next module. 
\end{itemize}


%
%
\begin{table}
\begin{adjustbox}{max width=\linewidth}
\centering
\begin{tabular}{c|r|c|c} 
\toprule
\multicolumn{2}{c}{\textbf{Model}}                                                                            & \multicolumn{1}{c}{\textbf{Baseline}} & \textbf{Proposed model}        \\ 
\toprule
\multirow{5}{*}{\rotatebox{90}{Encoder}} & Embedding                                                                & 512                                   & ~ ~ ~ 510                      \\ 
\cline{2-4}
                                   & \multirow{2}{*}{Classifier}                                              & –                                     & 2 $\times$ Bi-LSTM (510, 256)  \\
                                   &                                                                          & –                                     & FFN (256,2), Sigmoid           \\ 
\cline{2-4}
                                   & \multirow{2}{*}{\begin{tabular}[c]{@{}r@{}}Encoder \\core\end{tabular}}  & \multicolumn{2}{c}{3 $\times$ (512, 512)}                              \\
                                   &                                                                          & \multicolumn{2}{c}{Bi-LSTM (512, 256)}                                 \\ 
\toprule
\multicolumn{2}{c}{Attention}                                                                                 & \multicolumn{2}{c}{Location-sensitive (128)}                           \\ 
\toprule
\multirow{7}{*}{\rotatebox{90}{Decoder}} & \multirow{2}{*}{Pre-net}                                                 & \multicolumn{2}{c}{FFN (80, 256), ReLU}                                \\
                                   &                                                                          & \multicolumn{2}{c}{FFN (256, 256), ReLU}                               \\ 
\cline{2-4}
                                   & \multirow{2}{*}{\begin{tabular}[c]{@{}r@{}}Attention \\RNN\end{tabular}} & \multicolumn{2}{c}{LSTM (768, 1024)}                                   \\
                                   &                                                                          & \multicolumn{2}{c}{LSTM (1024, 1024)}                                  \\ 
\cline{2-4}
                                   & \multirow{3}{*}{Pst-net}                                                 & CNN (80, 512)                         & CNN(80, 510)                   \\
                                   &                                                                          & 4 $\times$ CNN (512, 512)             & 4 $\times$ CNN (512, 510)      \\
                                   &                                                                          & \multicolumn{2}{c}{CNN (512, 80)}                                      \\ 
\toprule
\multicolumn{2}{c}{Vocoder}                                                                                   & \multicolumn{2}{c}{Parallel WaveGan}                                   \\
\toprule
\end{tabular}
\end{adjustbox}
\caption{ Structure comparison between the baseline and the proposed method. }
\label{tbl:networkstructure}
\end{table}

\section{Experiments and Results}
After introducing the used corpus and discussing the training step, we define four evaluation Metrics. Finally we evaluate the methods through quantitative and subjective tests. \footnote{\textbf{Sound samples are available at}: \href{https://htmlpreview.github.io/?https://github.com/MahsaE/MahsaE.github.io/blob/master/Elyasi_IS21.html}{click-here}}
\label{sec:results}
\subsection{Corpus and Training}
We use a professional female American English speaker corpus, LJSpeech~\cite{LJSpeech}, that consist of approximately 24 hours of data. This corpus is commonly used for TTS training due to its size and public availability\footnote{Due to speaker characteristics, and utterance partition, not specifically at sentence break, synthesized speech may have lower quality compared with a system trained on higher-quality professionally private corpora}. 

For training, we normalize the text data and extract the prosodic-linguistic features using Festival~\cite{black1998festival}, and use Binary Cross Entropy loss to train our classifier. The classifier's parameters are learnt along with the rest of the TTM system in an end-to-end fashion. We use a training scheme similar to Shen et. al.~\cite{shen2018natural}'s, to train the TTM in Tacotron~2. We employ a pre-trained parallel waveGAN vocoder~\footnote{https://github.com/kan-bayashi/ParallelWaveGAN}~(trained on LJSpeech corpus). In our experiments, we trained five TTM models: TTM without any conditioning~(baseline), TTM with pre-encoder conditioning~(pre-encoder), TTM with pre-decoder conditioning~(pre-decoder), TTM with intra-decoder conditioning~(intra-decoder), and TTM with pre-encoder and intra-decoder conditioning~(combo). See Section~\ref{subsec:objective} for discussion on the \textit{combo} model. 

\subsection{Evaluation Metrics}
\label{subsec:eval_metrics}
We adapt two metrics that are commonly used for utterance similarity, with modified metrics, that provide better $F_{0}$ contour comparison between predicted signal and target signal: 
\begin{itemize}
    \item Mel Cepstral Distortion~(MCD)~\cite{kubichek1993mel} measures the difference~(MFCC) between two sequences of mel frequency cepstral coefficient. We use first 13 MFCC's and skip the first one~(overall energy).
    \item Utterance-level $F_{0}$ contour correlation~(UCorr) using  Pearson correlation methods that provides a number between $-1$ to $1$, where $0$ means no correlation, $1$ means total positive correlation, and $-1$ means total negative correlation.
    \item Partial Temporal $F_{0}$ contour correlation~(PTCorr) measures partial correlation between two signals at word-level with respect to ratio of word-duration to utterance-duration. Words with longer duration~(or prominent word) have more contribution than the words with shorter duration~(words like determiners or non-prominent words). 
    \item Partial $F_{0}$ contour variation measures partial $F_{0}$ range change at word-level in Hz.
\end{itemize}

\subsection{Quantitative Evaluation}\label{subsec:objective}
We randomly selected 50 fixed examples from the test set of the LJSpeech corpus as the evaluation set and present our findings in Table~\ref{tbl:Q_evl}. To account for the time-alignment mismatch between natural and synthesized speech for the MCD measure, we first apply dynamic time warping~(DTW) between MFCC of two speeches and then calculate the MCD between them. While, for the $F_{0}$ contour features~(rest of the Metrics), we use CMUSphinx toolkit~\cite{cmusphinx} to extract word alignment between the two signals. Then, we resample the $F_{0}$ contour within a word boundary of synthesized speech with respect to that of natural speech. \footnote{We note that although some studies first apply DTW, extract $F_{0}$, and then do alignment, extracting $F_{0}$ contour after DTW may results in an inaccurate (noisy) $F_{0}$ contour.}

Results of UCorr, shown in Table~\ref{tbl:Q_evl}, does not reveal any differences between the methods. This suggest that a globally correlation calculation can not fully represent how two $F_{0}$ contours are correlated. However, results of PTCorr bring out the differences between the methods. We see a PTCorr improvement of pre-encoder and intra-decoder, over the baseline. Results of partial $F_{0}$ variation show that intra-decoder method has relatively more variated $F_{0}$ dynamics than pre-encoder, pre-decoder, and the baseline. Separately conditioning at pre-encoder, pre-decoder, and intra-decoder do not improve the MCD measure over the baseline.

From the analysis of MCD, PTCorr, and partial $F_{0}$ variations, we conclude that pre-decoder only conditioning does not provide an improvement over the baseline, while pre-encoder and intra-decoder conditioning shows an improvement over the baseline for $F_{0}$ contour feature. After listening to the generated samples by the three methods we subjectively observe that: pre-encoder samples sound prosodically more smoother than the baseline, pre-decoder samples sound like ``the speaker has a cold'', and intra-decoder samples sound prosodically more valid than the baseline, but less smooth. Due to these observations, we combine pre-encode and intra-decoder as \textit{combo} method. Table~\ref{tbl:Q_evl}, last row, shows that the proposed \textit{combo} method not only results in higher PTCorr, and lower MCD, but also is closer w.r.t $F_{0}$ dynamics to natural speech.

\subsection{Subjective Evaluation: naturalness}
\label{ssec:subjective}

We use Amazon Mechanical Turk~\cite{crowston2012amazon}, and employ 150 turkers with approval ratings of at least $95\%$, located in United States. We run the following tasks: (a) to measure naturalness of recorded speech from LJSpeech corpus, (b) to measure naturalness of the baseline, and (c) to evaluate performance of the proposed method. For each task, we ask subjects to listen to an utterance, and judge the naturalness of the utterance on a five-point scale: 1 (bad), 2 (poor), 3 (fair), 4 (good), and 5 (excellent). The experiment includes 50 utterances for each of the three tasks that judged by 50 turkers. Three control utterances that were trivial to judge, were added to the experiment to filter out unreliable turkers, and each turker judged only utterances from one task. 

We randomly select 50 fixed examples from the test set of
the LJSpeech corpus as the evaluation set. Table~\ref{tbl:MOS} shows a comparison of our method against baseline and ground truth~(recorded speech). The results show that our method outperforms the baseline, and lead to a comparable MOS to that of the ground truth.

%
%
\begin{table}
\begin{adjustbox}{max width=\linewidth}
\centering
\begin{tabular}{rcccc} 
\toprule
\multirow{2}{*}{Method} & \multirow{2}{*}{MCD} & \multirow{2}{*}{UCorr} & \multirow{2}{*}{PTCorr} & Partial $F_{0}$ variation  \\
                        &                      &                        &                         & (Ground truth 48.26)       \\ 
\toprule
Baseline                & 5.36                     & 0.52                   & 0.60                    & 41.23                      \\
Pre-encoder             & 5.44                     & \textbf{0.55}                   & 0.62                    & 40.82                      \\
Pre-decoder             & 5.89                     & 0.51                   & 0.56                    & 40.78                      \\
Intra-decoder           & 5.60                     & 0.53                   & 0.67                    & 44.11                      \\
Combo                   & \textbf{5.09}                     & 0.54                   & \textbf{0.76}                    & \textbf{46.62}                      \\
\toprule
\end{tabular}
\end{adjustbox}
\caption{A summary of quantitative metrics (Section~\ref{subsec:eval_metrics}) used to evaluate all methods compared to ground truth (natural recording). Lower is better for MCD and higher for the UCorr and PTCorr.}
\label{tbl:Q_evl}
\end{table}

\begin{table}
\centering
\begin{adjustbox}{max width=\linewidth}
\begin{tabular}{rc} 
\toprule
Method         & MOS              \\ 
\toprule
Baseline       & 3.918$\pm$0.223  \\
Ground truth   & 4.288$\pm$0.101  \\
Proposed method & \textbf{4.148$\pm$0.192}  \\
\bottomrule
\end{tabular}
\end{adjustbox}
\caption{Mean Opinion Score (MOS) evaluations with $95\%$ confidence intervals.}
\label{tbl:MOS}
\end{table}
\section{Discussion and Future Work}
\label{sec:conclusion}
In this study we demonstrated a novel carefully designed strategy for conditioning Tacotron-2's TTM on prosodic-lingustic features. We conditioned Tacotron-2's TTM (baseline) at three different places~(see Section~\ref{sec:method} for explanation): Pre-encoder, Pre-decoder, and Intra-decoder. 

A set of objective measures was used to evaluate the performance of the synthesis models. Our quantitative experiment showed that conditioning at these three places individually does not result in a major metric improvement over the baseline. However, this observation lead us to combine Pre-encoder and Intra-decoder methods as a \textit{combo} method that not only resulted in higher PTCorr, and lower MCD, but also was closer w.r.t $F_{0}$ dynamics to natural speech. Through a set of subjective tests, we evaluated naturalness of the \textit{combo} method against the baseline and the natural speech. 

In the future, we want to investigate, how intentionally changing the prosodic-linguistic features at pre-encoder and intra-decoder would impact synthesized speech. For example, would it be interesting to investigate how ``zeroing-out'' one feature at-the-time in a specific place change the synthesized $F_{0}$ contour. In our approach we employ Festival for feature extraction, in the future, we would investigate methods to incorporate this into the proposed end-to-end conditioning strategy.

\bibliographystyle{IEEEtran}
\bibliography{0.main}

\begin{thebibliography}{10}
\providecommand{\url}[1]{#1}
\csname url@samestyle\endcsname
\providecommand{\newblock}{\relax}
\providecommand{\bibinfo}[2]{#2}
\providecommand{\BIBentrySTDinterwordspacing}{\spaceskip=0pt\relax}
\providecommand{\BIBentryALTinterwordstretchfactor}{4}
\providecommand{\BIBentryALTinterwordspacing}{\spaceskip=\fontdimen2\font plus
\BIBentryALTinterwordstretchfactor\fontdimen3\font minus
  \fontdimen4\font\relax}
\providecommand{\BIBforeignlanguage}[2]{{%
\expandafter\ifx\csname l@#1\endcsname\relax
\typeout{** WARNING: IEEEtran.bst: No hyphenation pattern has been}%
\typeout{** loaded for the language `#1'. Using the pattern for}%
\typeout{** the default language instead.}%
\else
\language=\csname l@#1\endcsname
\fi
#2}}
\providecommand{\BIBdecl}{\relax}
\BIBdecl

\bibitem{yasuda2020investigation}
Y.~Yasuda, X.~Wang, and J.~Yamagishi, ``Investigation of learning abilities on
  linguistic features in sequence-to-sequence text-to-speech synthesis,''
  \emph{Computer Speech \& Language}, p. 101183, 2020.

\bibitem{shen2018natural}
J.~Shen, R.~Pang, R.~J. Weiss, M.~Schuster, N.~Jaitly, Z.~Yang, Z.~Chen,
  Y.~Zhang, Y.~Wang, R.~Skerrv-Ryan \emph{et~al.}, ``Natural tts synthesis by
  conditioning wavenet on mel spectrogram predictions,'' in \emph{2018 IEEE
  International Conference on Acoustics, Speech and Signal Processing
  (ICASSP)}.\hskip 1em plus 0.5em minus 0.4em\relax IEEE, 2018, pp. 4779--4783.

\bibitem{wu2016merlin}
Z.~Wu, O.~Watts, and S.~King, ``Merlin: An open source neural network speech
  synthesis system.'' in \emph{SSW}, 2016, pp. 202--207.

\bibitem{black1998festival}
A.~Black, P.~Taylor, R.~Caley, and R.~Clark, ``The festival speech synthesis
  system,'' 1998.

\bibitem{watts2019improvements}
O.~Watts, G.~E. Henter, J.~Fong, and C.~Valentini-Botinhao, ``Where do the
  improvements come from in sequence-to-sequence neural tts?'' in \emph{2019
  ISCA Speech Synthesis Workshop (SSW)}, vol.~10, 2019, pp. 217--222.

\bibitem{liu2019cross}
Z.~Liu and B.~Mak, ``Cross-lingual multi-speaker text-to-speech synthesis for
  voice cloning without using parallel corpus for unseen speakers,''
  \emph{arXiv preprint arXiv:1911.11601}, 2019.

\bibitem{liu2020multi}
------, ``Multi-lingual multi-speaker text-to-speech synthesis for voice
  cloning with online speaker enrollment,'' \emph{Proc. Interspeech 2020}, pp.
  2932--2936, 2020.

\bibitem{fujimoto2019impacts}
T.~Fujimoto, K.~Hashimoto, K.~Oura, Y.~Nankaku, and K.~Tokuda, ``Impacts of
  input linguistic feature representation on japanese end-to-end speech
  synthesis,'' in \emph{10th ISCA Speech Synthesis Workshop. ISCA, Vienna,
  Austria}, 2019.

\bibitem{yasuda2019investigation}
Y.~Yasuda, X.~Wang, S.~Takaki, and J.~Yamagishi, ``Investigation of enhanced
  tacotron text-to-speech synthesis systems with self-attention for pitch
  accent language,'' in \emph{ICASSP 2019-2019 IEEE International Conference on
  Acoustics, Speech and Signal Processing (ICASSP)}.\hskip 1em plus 0.5em minus
  0.4em\relax IEEE, 2019, pp. 6905--6909.

\bibitem{luong2018investigating}
H.-T. Luong, X.~Wang, J.~Yamagishi, and N.~Nishizawa, ``Investigating accuracy
  of pitch-accent annotations in neural network-based speech synthesis and
  denoising effects,'' \emph{arXiv preprint arXiv:1808.00665}, 2018.

\bibitem{zen2019libritts}
H.~Zen, V.~Dang, R.~Clark, Y.~Zhang, R.~J. Weiss, Y.~Jia, Z.~Chen, and Y.~Wu,
  ``Libritts: A corpus derived from librispeech for text-to-speech,''
  \emph{arXiv preprint arXiv:1904.02882}, 2019.

\bibitem{taylor2019analysis}
J.~Taylor and K.~Richmond, ``Analysis of pronunciation learning in end-to-end
  speech synthesis.'' in \emph{INTERSPEECH}, 2019, pp. 2070--2074.

\bibitem{liu2020tone}
R.~Liu, X.~Wen, C.~Lu, and X.~Chen, ``Tone learning in low-resource bilingual
  tts,'' \emph{Proc. Interspeech 2020}, pp. 2952--2956, 2020.

\bibitem{suni2020prosodic}
A.~Suni, S.~Kakouros, M.~Vainio, and J.~{\v{S}}imko, ``Prosodic prominence and
  boundaries in sequence-to-sequence speech synthesis,'' \emph{arXiv preprint
  arXiv:2006.15967}, 2020.

\bibitem{wang2017tacotron}
Y.~Wang, R.~Skerry-Ryan, D.~Stanton, Y.~Wu, R.~J. Weiss, N.~Jaitly, Z.~Yang,
  Y.~Xiao, Z.~Chen, S.~Bengio \emph{et~al.}, ``Tacotron: Towards end-to-end
  speech synthesis,'' \emph{arXiv preprint arXiv:1703.10135}, 2017.

\bibitem{yamamoto2020parallel}
R.~Yamamoto, E.~Song, and J.-M. Kim, ``Parallel wavegan: A fast waveform
  generation model based on generative adversarial networks with
  multi-resolution spectrogram,'' in \emph{ICASSP 2020-2020 IEEE International
  Conference on Acoustics, Speech and Signal Processing (ICASSP)}.\hskip 1em
  plus 0.5em minus 0.4em\relax IEEE, 2020, pp. 6199--6203.

\bibitem{karras2019style}
T.~Karras, S.~Laine, and T.~Aila, ``A style-based generator architecture for
  generative adversarial networks,'' in \emph{Proceedings of the IEEE/CVF
  Conference on Computer Vision and Pattern Recognition}, 2019, pp. 4401--4410.

\bibitem{liu2018intriguing}
R.~Liu, J.~Lehman, P.~Molino, F.~P. Such, E.~Frank, A.~Sergeev, and
  J.~Yosinski, ``An intriguing failing of convolutional neural networks and the
  coordconv solution,'' \emph{arXiv preprint arXiv:1807.03247}, 2018.

\bibitem{LJSpeech}
\BIBentryALTinterwordspacing
K.~Ito, ``The lj speech dataset.'' [Online]. Available:
  \url{https://keithito.com/LJ-Speech-Dataset/}
\BIBentrySTDinterwordspacing

\bibitem{kubichek1993mel}
R.~Kubichek, ``Mel-cepstral distance measure for objective speech quality
  assessment,'' in \emph{Proceedings of IEEE Pacific Rim Conference on
  Communications Computers and Signal Processing}, vol.~1.\hskip 1em plus 0.5em
  minus 0.4em\relax IEEE, 1993, pp. 125--128.

\bibitem{cmusphinx}
\BIBentryALTinterwordspacing
``Cmu sphinx-4: open source speech recognition toolkit.'' [Online]. Available:
  \url{http://www.speech.cs.cmu.edu/sphinx/}
\BIBentrySTDinterwordspacing

\bibitem{crowston2012amazon}
K.~Crowston, ``Amazon mechanical turk: A research tool for organizations and
  information systems scholars,'' in \emph{Shaping the future of ict research.
  methods and approaches}.\hskip 1em plus 0.5em minus 0.4em\relax Springer,
  2012, pp. 210--221.

\end{thebibliography}
\end{document}